# Throwing Out the Baby with the Bathwater: The Undesirable Effects of National Research Assessment Exercises on Research

**John Mingers**

Kent Business School, University of Kent, Canterbury, UK,

P: 01227 824008

e: j.mingers@kent.ac.uk

**Leroy White**

Warwick Business School, University of Warwick, Coventry UK

e: Leroy.White@wbs.ac.uk

**Abstract**:

The evaluation of the quality of research at a national level has become increasingly common. The UK has been at the forefront of this trend having undertaken many assessments since 1986, the latest being the "Research Excellence Framework" in 2014. The argument of this paper is that, whatever the intended results in terms of evaluating and improving research, there have been many, presumably unintended, results that are highly undesirable for research and the university community more generally. We situate our analysis using Bourdieu's theory of cultural reproduction and then focus on the peculiarities of the 2008 RAE and the 2014 REF the rules of which allowed for, and indeed encouraged, significant game-playing on the part of striving universities. We conclude with practical recommendations to maintain the general intention of research assessment without the undesirable side-effects.

# INTRODUCTION[1]

> "People know what they do; frequently they know why they do what they do; but what they don't know is what what they do does." Michel Foucault quoted in Dreyfus (1982, p. 187)

It is not an exaggeration to say that the single factor that has most driven behaviour within the university sector over the last twenty years in the UK is the Government's regime of research assessment. It is a powerful lever of control over the academic community (Broadbent, 2010). This began in a limited way in 1986 with the Research Selectivity Assessment (RSA)[2], which was designed to help the Government distribute its research funds more effectively. Universities did not pay it that much attention. However, as successive assessments unrolled, and it became clear that the results could be used to create research league tables (primarily in the *Times Higher* newspaper) and that these league tables had a huge effect on the fortunes of universities, success in the research assessment has become the number one priority for the majority where research plays a significant role. The amount of time spent both in universities managing the processes and in panels assessing the research is truly huge and to the extent that it is undertaken by academics is itself a diversion from actually doing research.

In later research assessments, the proclaimed objective has moved from simply evaluating research quality to supposedly improving the standing of the UK as a whole, both in terms of quantity but more importantly quality. However, many academics feel that, whatever the effects on the quantity and quality, the RAs have also had many, presumably unintended, negative effects, both in terms of research specifically, and in terms of the culture of the university sector as a whole. Perhaps, in trying to eliminate the weaker "tail" of poor quality research, the baby of high quality, truly innovative, practically oriented research has been thrown out as well?

There are two particular issues that provide a context for this paper. The first issue is the differentiation in the field of British universities that often distinguishes the traditional elite universities, such as the Russell Group, while simultaneously encouraging systemic expansion elsewhere. Studies show that this differentiation is not well accepted in that universities consistently strive to reposition themselves (Gonzales, 2014;O'Meara, 2007). Thus, universities tend to have similar visions to seek to become like those directly above them on the reputation hierarchy (such as league tables and the like) that is recognised within the UK higher education system. For decades now, these "striving" universities adopt complex mechanisms to achieve such aspirations. The strategies adopted for the RA exercise is one such mechanism. Here, advocates of the RA argue that as a mechanism it is the best (or least worst) way of assessing research quality in order to both improve performance and ensure the quality of the universities (Morphew and Baker, 2004).

In theory, individual universities select staff/outputs on the basis of quality, informed by publicly available performance information. In a system in which funding follows performance, universities devise strategies to improve that quality in order to attract more funding. If universities have the incentive to respond to improve their research quality – and all universities are equally able to do so – a research assessment-based system should not suffer an elite university 'bias' but support a striving attitude. The extent to which this is realized in practice depends both on the political context being considered and, more fundamentally, on the rules of the game (broadly defined) for both the universities and the authorities engaging in the process (Gonzales, 2014). It seems that striving is attributed to the universities' obsession with cultural resources such as reputation and legitimacy (Morphew, 2009), where it is suggested that the acquisition of cultural resources is inextricably intertwined with the need for universities to replace economic resources that government no longer provides.

The second issue is the RA context itself. The generally observed outcome, across different disciplines, is that universities' research quality has not improved. At the same time, questions have been raised concerning the equity of the model and possible deleterious effects on research in general (Becher and Trowler, 2001). Is it that practices intended merely to measures research productivity themselves create particular dynamics of power and produce or sustain particular hierarchies regarding types of research and models of knowledge production (Alldred and Miller, 2007)?

At the heart of these issues is the process of the RA itself and the mechanisms employed by individual universities in their interactions with this method of assessing research quality. How and why universities choose between staff and outputs – and the extent to which they are able to realize their research quality – has a fundamental impact on the outcomes of RA mechanisms for the different universities, and raises subsequent questions regarding policy design. Addressing this question in the context of RA is the focus of this paper

In this paper, we draw on the work of Bourdieu to provide an analytical framework that we hope will prove to be illuminating on the issue of strategies and consequences. In particular, we focus on his cultural reproduction theory which has come to dominate the sociological literature on education and to some extent universities (Bourdieu, 1984, 1986;Bourdieu and Passeron, 1977). We note in passing that Bourdieu himself did not systematically analyze performances of universities. But his work, and concerns about the increasing dominance of marketization and state regulation of both knowledge production and pedagogic transmissions, led him to develop the outlines of an analytical framework that has much to offer for the analysis of change within the universities. We apply this theoretical perspective on the RA in the context of selectivity for the RA exercise in order to better understand first, differences in the mechanisms of selectivity; second, the ways in which such differences explain the observed sometimes unintended outcomes; and, third, how the institutional culture created by the performance management framework within which universities operate may impact on those differences. We show that the strategies adopted or created by such a system may serve to exacerbate rather than alleviate issues relating to research quality. We argue that situating RA within an institutional context provides a coherent and parsimonious framework in which to compare intended and unintended effects of RA decision making and to analyze the effects of policy recommendations.

In this paper we will identify many changes that have taken place in the research community over the last twenty years and show how many of them are directly the result of successive RAs and the particular, often quite bizarre, rules that govern them. After some general comments, we will focus on the business and management (B&M) subject area as that is where our experience lies, although it is also the biggest single subject area of the whole assessment.

## THEORETICAL PERSPECTIVE: FIELD, HABITUS AND CULTURAL CAPITAL

There are a growing number of studies that offer different explanations of the mechanism and consequences of the RA exercise (Macdonald and Kam, 2007;Alldred and Miller, 2007;Mingers and Willmott, 2013). There is also a growing research tradition that draws on Bourdieu showing that there are differences in universities which reflect wider persistent education inequalities despite universities expansion (Gonzales, 2014).

Most of the explanations for the strategies for the RA, we argue, can be seen in light of the work of Bourdieu. Bourdieu (2005) mentions the organisation as fields and speaks of educational institutions as fields. Briefly, a field is defined as a configuration of positions comprising agents (individuals, groups of actors or institutions) struggling to maximise their positions. Agents are defined by their relational position within a field's distribution of capital. The structure of the field is governed by relations between these positions. Central to Bourdieu is the dynamic of domination and the relation between dominator and dominated. The field is, then, a site of ongoing struggle: structures of power reproduce and are reproduced by inequality. The field is also a structured hierarchy in the sense that agents and institutions occupy dominant and subordinate positions, which depends on the amount of capital that are possessed. Thus, universities are conceptualised as a field with a high degree of autonomy in that the field generates its own values and behavioural imperatives which are relatively independent from forces emerging from economic and political fields.

As well as the dominance of certain values across society as a whole, within universities there are differences in the amounts and mixes of different forms of capital (Bourdieu, 1986). These include not just economic capital but also social capital (the power and resources available from social networks) and particularly cultural capital. In thinking about the role of universities, Bourdieu wanted to get away from human capital theories (that emphasise the economic returns to education). Rather he was interested in identifying a form of capital which is not so easily converted to money but that had a strong effect on differences individuals and entities: cultural capital.

Cultural capital comprises the knowledge, skills, taste, preferences and possessions that give advantages (or disadvantages) in the system of relations (Bourdieu, 1986). These aspects can take different forms, from aesthetics, tastes, and ways of speaking acquired through socialization, to the institutionalized knowledge and skills acquired through formal training and qualification. Once acquired these institutionalised forms of cultural capital have symbolic potential to convey issues of reputation. Cultural capital is in three forms: embodied, institutionalised and objective. Embodied cultural capital is literally inscribed on the body – in bearing and presence, styles of conduct, habits and demonstrations of taste – over for example food, furniture or art. Institutionalised cultural capital exists in the form of qualifications and credentials. Objective cultural capital exists in objects and material resources – articles, books etc. In Bourdieu's work institutions are distinguished by different endowments of economic, social and cultural capital (Bourdieu, 1984). The elite universities have higher amounts of capital overall but in particular (especially in relation to the research field) higher cultural capital. These forms of cultural capital are also the ones valorised by society as a whole such that, for example, elite universities' displays of taste are seen as legitimate 'good taste'. Cultural capital takes time to acquire and is not easily lost, nor easily bought by economic capital. Its acquisition and duration make it a strong element in elite university reproduction and university distinction. It is also the element that is most strongly interconnected with institutionalised forms of educational provision in societies. It is for this reason that research into consequences of RA, from this Bourdieusian perspective, will focus on cultural capital, its transmission, acquisition and deployment as the key mechanism for exploring differences in strategies over RA and differences in performance more widely.

Bourdieu's form of cultural capital theory is operationalised here as cultural capital being a form of comparative advantage for universities in the higher education market. We explore whether cultural capital is the dominant mechanism to account for differences in strategies for the RA. Cultural capital is acquired over time and the main mechanism of transmission is via university striving through acquiring cultural capital – so some universities have a comparative advantage in the system and come from environments with more objective cultural capital (more educational resources). We introduce the notion of "striving" here to underscore the characteristics attributed to all universities as almost a dimensional quality. Thus, striving is not a single, discrete attribute but one where we assume that universities are more or less striving depending on their cultural capital. Equally more striving universities themselves have more cultural capital than less striving ones. This involves the different forms of cultural capital – embodied, institutionalised and objective. They have more knowledge of the system overall and ways of addressing university professionals that gets the best out of them. This cultural capital based explanation does not assume that more or less striving universities have different values towards research quality. Rather some universities have more cultural capital than others; they are able to transmit this to staff; and their staff are able to deploy cultural capital in the RA context: i.e. knowledge of the system is part of the cultural capital that immediately separates out many universities and orients them differently to RA decision making. These elements combine with the fact that these cultural deployments are closer to the culture of the universities that then enhances those deployments. As well as being endowed with larger and more effective amounts of cultural capital more striving universities are also closer to the dominant culture whereas less striving universities are at a further distance from that culture.

Distilling the various arguments in this approach to RA and research quality seems to imply that the different endowments of cultural capital between the universities directly affect the mechanism of the strategies adopted. The different strategies can be explained by the effects of cultural capital in opening the space for greater investment in the process and having more to gain from the outcome of it for striving universities. As Bourdieu might argue, it is not just the deployment of cultural capital that is significant but the space for investment and reward it creates both going into the situation and emerging from it.

## THE EVOLUTION OF RESEARCH ASSESSMENT IN THE UK

The process of formal research assessment in the UK began in 1986 with what was then known as the research selectivity exercise (RSA). It occurred again in 1989, 1992, 1996 (now called the Research Assessment Exercise, RAE), 2001, 2008 and 2014 (Research Excellence Framework, REF). See Appendix 1 for a summary of the main features of each one. The primary purpose has always been to assist the Government in handing out its block research funding (known as "QR" money) between the different university departments by assessing their "relative quality". But it has also been seen as a way of improving the quality of UK research as stated in the Roberts review after the 2001 RAE:

*"The system was designed to maintain and develop the strength and international competitiveness of the research base in UK institutions, and to promote high quality in institutions conducting the best research and receiving the largest proportion of grant." (Roberts, 2003, p. 1)*

The exercise has changed in many ways over the years, often in response to reviews and also a good deal of criticism from academics (see Bence and Oppenheim (2005) for an overview) but in this section we will simply highlight the main characteristics that have been contentious and have potentially had negative impacts. Several reviews of the business and management assessments have been written by the panel members after the event although, in our view, they tend to be somewhat vapid (Ashton et al., 2009; Bessant et al., 2003; Cooper and Otley, 1998).

**a. Selectivity of staff**. One of the major characteristics of all the RAs has been the decision to allow selectivity in the number of staff that are submitted. This immediately offers the opportunity for universities to choose only the more research active staff and thereby improve their score. As Roberts says, "This is a source of some controversy. Institutions can, by excluding their weaker researchers, obtain higher grades than they would otherwise do" (Roberts, 2003, p. 34). In 1992 a letter grade indicating the proportion of staff being submitted was added to the score but this was not reflected in the league tables. It was later dropped. In 2008 HEFCE had intended to present the proportion of staff included in the results but after the threat of legal action by the Russell Group of universities (who would presumably have suffered from this), this also was dropped. It is intended that the proportion will be available for the 2014 exercise.

Roberts considered the possibility of including all staff in a department but concluded "*there is a real risk that institutions would respond with an even more damaging form of games-playing – removing reference to researchers from the contracts of large numbers of teacher-researchers*" (Roberts, 2003, p. 34). He recommended that departments should include at least 80% of their staff but this was not taken up by HEFCE. In 2008, in business and management, most (but significantly not all) submissions did try to include a high proportion of staff, but the success of those who did not has led to a much greater degree of selectivity in 2014 . Indeed, it is now the case that the decision about how selective to be (requiring a GPA of 2.75, 3.0 or even 3.5, see below) is likely to have a greater effect on the final results and league tables than all the research that is actually carried out and documented. As has always been the case, many of the rules of the game are not actually known at the time of submission. In this case, will the league tables produced simply record the grade point average, or will they be adjusted in some way for the proportion or absolute number of staff entered? If the former, then the more selective the entry the better the score; if the latter, then those being more inclusive should gain. But surely something as important as the REF should not be reduced to a lottery in which those who best guess the rules win.

In the event in 2014 the final *Times Higher* table was ordered in terms of GPA but also showed research "power", that is, the GPA multiplied by the number of researchers submitted. This reveals the extent to which institutions were inclusive in their submissions. For example, Cardiff Metropolitan University was placed 41$^{st}$ – an amazing result for a non-research intensive university. However, only 35 staff were submitted from the whole university! On power it comes 95$^{th}$. LSE came 3$^{rd}$ overall on GPA but would have been 28$^{th}$ on power. On the other hand, Nottingham only came 26$^{th}$ but would have been 7$^{th}$ on power.

HESA produced statistics on the number of staff who were eligible to be submitted by each departmental submission. Strangely, these were not published until the 18$^{th}$ December, the day the Times Higher ranking was published. As the Times Higher itself says (page 34), the data was not published (although obviously available) in time for it to be incorporated in the tables. One has to question why that was.

With this data, it is possible to calculate the percentage of staff entered for each submission. In business and management, some major departments had relatively small submissions, for example Aston (43%), Cardiff (56%), Oxford (51%), Reading (61%), Sheffield (55%), Surrey (56%). The first five of these were all in the top 20 on GPA.

So, in fact, we have three totally different scores available – the GPA, the GPA weighted by the number submitted (power) and the GPA weighted by percentage submitted (intensity). Indeed, the Association of Business Schools (ABS) has also produced a "table of tables". The university of one of the authors' potentially has three very different ranks – $41^{st}$ (GPA), $33^{rd}$ (power), and $25^{th}$ (intensity). No prizes for guessing which will be used. There are, in fact, many other possibilities - % 4*, %( 4*+3*), impact or environment, each of which can be weighted by intensity or number submitted! This really makes a nonsense of the whole process.

There is also evidence that the number of staff submitted was illegitimately affected by the number of impact case studies that a department had. One case study was needed for every ten staff submitted but if there was a shortage of cases the number of submitted staff was reduced.

**b. Selectivity of outputs**. As with staff, only a proportion of research outputs need be submitted. In fact, for the last few RAs the number was restricted to a maximum of four per person submitted. The primary reason for this is believed to be simply workload – to have all research submitted would be too much for the Panels. HEFCE's view was that the RA was only intended to assess top level research, not all research. This leads to problems over choosing which outputs to select.

**c. The grading system**. Until 2008, the grading system awarded each submission a single grade although the scales differed. The exact manner in which the Panel came to its decision, based on the evidence in the submission, was never really very transparent. Moreover, such a crude grading system had serious problems, as was highlighted by Roberts. This method was changed dramatically for 2008 – each submission was to be given a quality profile across five grades (0 – 4*), where 0 represented "not research" and 4* represented "world leading research", showing the proportion of work judged to be in each category. The profile was generated from three components, the outputs, the research environment and esteem factors. The weights for these varied across the subject. The major innovation here was that it required that every single output submitted had to be given its own grade. Apart from a huge amount of work, this this created many problems for departments trying to decide which outputs (and therefore staff) to submit. It also significantly increased the reward for game-playing – submitting only those staff who were judged to have high quality outputs (3* and 4*) would directly affect the quality score. The league table compilers simply averaged the profile scores to give a "GPA", typically in the range 2.0 to 3.3.

In 2008 the mean GPA across all submissions in all subjects was xxxxx. In 2014 it had risen to 3.01, a very significant rise. Does this mean that the actual quality of research has risen by this amount, or is it simply grade inflation? This 3.01 GPA is actually higher than the score for the University of Cambridge in 2008 – does that mean that the average university is now better than Cambridge was only five years ago?

**d. Differential funding**. The main purpose of the RA was to allocate funding, but clearly there were never sufficient funds to support all the research. HEFCE therefore had to be cautious and generally only decided how it would allocate *after* the results were announced. So effectively institutions were playing a game in which the main rule was only announced after the score was in. In 2008, funding was heavily skewed towards 3* and 4*, and in 2014 it is believed it will be almost exclusively 4* putting ever greater emphasis on only work judged to be of the highest quality (generally this is seen as papers published in journals considered to be 4*).

**e. Interdisciplinary, multidisciplinary and practice based research**. The rhetoric of the RAs has always been that it encourages these non-traditional forms of research and that all forms of research outputs should be equally valued. However, the reality has been very different as was recognised by the Roberts report (and see, for example Lee (2013) for a discussion of this in Economics). In practice, as we will discuss later, the emphasis has swung hugely towards theoretical papers published in the top (single) discipline journals.

One of the biggest changes in recent years was the decision to include non-academic impact as one of the factors to be assessed in 2014. This counted for 20% of the overall score, a very significant amount, and is to be assessed by the submission of case studies demonstrating the impact of research. This in itself has caused huge problems especially for newer departments, or departments that have a significant proportion of new staff since impact does not transfer with a member of staff whilst, bizarrely, the research outputs do. Whilst this clearly addresses the issue of practice-based research in a significant way quite how it will work in practice remains to be seen.

Following Bourdieu, we can determine the specific form of power, which individual universities and staff possess in terms of the accumulation of research capital. Examples of capital include qualified staff, outputs grading and research funding. This leads to a large variation in the distribution of this form of capital. This institutionalised form of capital is the context in which there is pressure to play the publication game and strive to submit to journals of high international standing as well as pressure raise international standing of universities. Research selectivity is determined on the judgement on those researchers who, as Bourdieu maintained, invest above all in production and also in the work of representation which contributes to the accumulation of cultural capital of external recognition or renown. Given the importance of selectivity for the RA, activities to boost the external recognition are all but evident.

It seems clear that cultural capital, as well as the material benefits derived from funding that follows from the RA exercise, are considered highly important, at the very least by the administrative powers and serves to inform the direction of the research being done within universities. From Bourdieu's field perspective, universities are nothing but a secular battleground filled with capital market practice such as playing the power game, competing for resources, and mechanism conversion from one type of capital to another. In particular, the conflict between academic and administrative power takes centre stage in this battle. In the university field, the above mentioned power game refers to not only the struggle for legitimacy between academic power founded on professional disciplines and administrative power inseparable from political capital and complying with bureaucracy norms, but also the reproduction of their legitimacy. Such struggles for legitimacy between academic and administrative power is clearly demonstrated in the RA strategy adopted, the conflict between administrative staff and university research staff, as well as compromising and co-existing strategies for their respective development. The reproduction mechanism of their legitimacy, on the other hand, follows separated and independent modes of development. In this way, the open and integrated co-existing system within universities is strengthened so as to ensure the vitality and reproduction of academic and administrative power. But despite the fact that less-striving universities are seriously disadvantaged in the competition for research credentials, the results of this competition are seen as meritocratic and therefore as legitimate. In addition, in Bourdieusian terms, inequalities are legitimated by the research credentials generated through the RA and held by those in dominant positions. This means that the RA system has a key role in maintaining the status quo.

## HAVE THE ASSESSMENTS IMPROVED THE QUALITY OR QUANTITY OF RESEARCH?

Although the original aim of the research assessment was to decide how to allocate Government money, later assessments were also justified on the grounds that they would improve the quantity and/or quality of UK research. Has this in fact happened?

We will begin by considering UK performance as a whole as some data is available for that. Adams (2011), in a report on UK research for Thompson Reuters (who own Web of Science), looked at comparative UK performance from 1991 to 2008. The main conclusions were:

1. Whilst the number of UK papers has risen significantly (from 50,000 to 90,000) so have those from other countries, especially the new economies such as China, so that after a rise our share of world publications is actually now falling and is around 8% - 9% (see Adams, Figure 4).

2. Our impact, as measured by citations, is rising and our normalised citation impact is now ahead of the US, as well as Germany and France (see Adams, Figure 6).

3. We also have more than our share of exceptionally highly cited papers, for example, between 2000 and 2010 we had 603 that were cited over 500 times (which represented 16.7% of the total) and 142 cited over 1000 times (19.8% of the total) in comparison with the proportion of papers that is 9% (see Adams, Table 2).

Whilst this appears good it is almost entirely based on science rather than social science, let alone business and management. The areas where impact is greatest are biological sciences, astronomy and astrophysics, and geosciences (Thompson Reuters, 2014). As an illustration of this, we looked up in WoS the papers published with a GB address, which had more than 1000 citations between 2000 and 2010 – there were 342. Of these, only 2 were for non-science subjects and they were both in economics[3].

We will now look more specifically at the business and management field. Figure 1 shows the results of a search of WoS for the years 2000-2012 which covers the latest three research assessments. The top line shows the total number of papers published in English in the WoS categories "Business" or "Management". The bottom line shows the same search but restricted to Country (in the address field) as "England".

------------------------------------------
INSERT FIGURE 1 ABOUT HERE

------------------------------------------

These show a doubling of the number of papers published over the period (11,000 to 21,600 for the total, and 900 to 2200 for England). The corresponding proportion of England papers is shown in the top line in Figure 2 – it has risen from around 8% to around 10%, so there appears to have been a rise in relative quantity. It is interesting to note that there are peaks in the proportion in the years 2000, 2007 and 2012, the years of RA submission. Moed (2008) also detected significant patterns of response to research assessment metrics, with an increase in total publications after 1992 when numbers of papers were required; a shift to journals with higher citations after 1996 when quality was emphasised; and then in increase in the apparent number of research active staff through greater collaboration during 1997-2000.

Is there also an improvement in quality? To the extent that quality can be measured by citations, we would expect this to show in an increase in the most highly cited papers. The generally accepted level for a highly cited paper is that it is in the top percentile (see for example *Essential Science Indicators* which use this metric), i.e., the most highly cited 1% of papers.

------------------------------------------
INSERT FIGURE 2 ABOUT HERE

------------------------------------------

To measure this we have found the number of citations received by the top percentile paper in the total set (say this is 155 citations). We then find how many papers from the England subset have received this many citations and calculate what percentage this is of the total

England papers. If the result is close to 1% then the England papers have a similar performance to the total set. If it is greater than 1% then they are over-performing. If quality is increasing, we would expect this percentage to be rising. We can see from the bottom line that the proportion is almost flat, the regression has a coefficient of 0.02 which is statistically insignificant.

So the conclusion from this section is that the quantity of papers has increased both in absolute terms and as a proportion of the world total, however there is no evidence that the impact of research, at least as measured by the proportion of very highly papers, has improved.

# THE INTENDED AND UNINTENDED CONSEQUENCES OF THE RESEARCH ASSESSMENT

In this section we will delineate some of the main negative consequences of successive research assessments. We will be discussing the business and management (B&M) area specifically, although many of these effects apply in other disciplines. We will also limit ourselves to the two latest assessments (2008 and 2014) as these involved significant changes to the grading system which we believe exacerbated the problems.

The causes and consequences are summarised in the influence diagram in Figure 3. In this diagram, the rectangular boxes show the main characteristics of the RAE/REF as discussed in Section 3 which, we argue, have had undesirable effects on research. The rounded boxes show what those consequences are, and the factors not in boxes are mediating influences. The various causal chains depicted in this diagram are explained in details in the next sections.

-------------------------------------------
INSERT FIGURE 3 ABOUT HERE

-------------------------------------------

**The Grading System and the Taylorization of Research**

Bourdieu's view is that cultural capital is inculcated in the striving universities, and enables elite universities to gain higher research credentials than less-striving ones. This enables striving universities to maintain their position, and legitimates the dominant position which striving universities typically go on to hold. Of course, *some* less-striving universities will succeed in the RA system, but, rather than challenging the system, this will strengthen it by contributing to the appearance of meritocracy. In a system that purports to be strongly fair, rather than being purely based on the rationality of economic capital, differences in RA selection strategies can be explained by differences in cultural capital. What is suggested is that cultural capital does have a significant impact on RA outcomes and in maintaining the status quo. The field is the site of construction of strategies and struggles based on inherent interests and dispositions. The competing strategies and struggles for the accumulation of academic and research capital (embodied, institutionalized and objective) form the main focus of analysis.

One of the most profound problems for B&M revolves around the grading system and the evaluation of outputs. The 2008 exercise, for the first time, required panels to grade every single output – a huge task for them. This posed an even bigger problem for institutions when this was combined with the selectivity of outputs – having to choose which four to submit, and selectivity of staff – having to choose which staff to submit. This made it necessary for institutions to try to predict a score for every output produced by their staff. How was this to be done? Perhaps the obvious answer was to have them peer reviewed by experts either within or without the department. This was and is done, but there are problems – peer reviewing is ultimately subjective and there will be genuine differences of opinion (not to mention deliberate foul play); internal staff often do not wish to give a poor grade to their colleagues' work; there is no guarantee that it will give the same result as the REF panel; it is complex and time-consuming; it requires the involvement of a lot of people who may apply different standards; and it can lead to arguments and disputes.

At the same time, the emphasis of the B&M Panel was clearly on refereed journal papers (Stewart, 2005). Although the rhetoric of the REF said that all types of output should be treated equally, in practice it had become clear that refereed papers were seen as the primary commodity of the exercise. Indeed, this was made clear in verbal statements by the Panel Chair and members[4]. Certainly it was believed by institutions as the proportion of journal papers rose from 69% in 1996 to 80% in 2001 to an amazing 92% in 2008. Clearly books, book chapters and applied research reports are now almost valueless. This can also be seen in an analysis of the 2008 results (Mingers, Watson and Scaparra, 2012b) which estimated that papers received a mean score of 2.34, authored books 2.44, book chapters 2.21, edited books 2.10, and external reports 1.13. Whilst the score for books is marginally higher than for papers, one would have expected that a good research monograph should be worth considerably more than a single paper.

This meant that in practice what was necessary was a way of estimating the quality of journal papers. A second alternative would therefore be bibliometrics, particularly citations. Indeed, in the run-up to the 2008 exercise there were proposals that citations would be used in the exercise itself. However, testing showed that this was not a viable proposal and it was dropped. This diminished its utility in the eyes of decision makers such as Deans and Directors of Research (DoR). A third alternative was to judge the book by its cover and equate the quality of a paper to the quality of the journal it was published in. There were of course problems with this – not all papers within a journal are in fact of the same quality – you can get very good papers in a generally low quality or perhaps new journal and vice versa; and how do you judge the quality of a journal anyway? There were at the time several ranking lists of journals available (Mingers and Harzing, 2007), collected together on the Harzing website (Harzing, 2009), but they were to some extent incommensurable – created for different purposes; by different groups of academics; using different scales; and covering different sets of journals. It was at this point that the *Association of Business Schools* journal ranking list appeared (ABS list) as something of a saviour.

The ABS list began as a list compiled at Bristol Business School, "not intended for general circulation", based on journals submitted to RAE 2001 plus some others, with the grades standardised to 2008 RAE 1 - 4. Decisions were made by the editors, Huw Morris and Charles Harvey. The ABS, consisting of Deans and Directors of business schools, decided that it needed to produce a single, ranked list and adopted the Bristol one in 2007, augmenting it with input from subject specialists and journal impact factors. New versions were brought out in 2008, 2009, and 2010 (Association of Business Schools, 2010) with marginal revisions being made each time. The list quickly established itself as the *de facto* standard for judging the quality of papers in B&M and became used not only for decisions about the REF but also for promotions, hiring and guidance for academics as to where to publish. It is hard to over-estimate the influence of the ABS list in business research.

The list itself has received a huge amount of criticism both generally (Willmott, 2011;Mingers and Willmott, 2013;Tourish, 2011;Adler and Harzing, 2009;Ozbilgin, 2009) and from fields within business and management, for example accounting (Hussain, 2013, 2011;Hoepner and Unerman, 2009), operational research (Mingers et al., 2012b) (the Committee of Professors in OR reluctantly produced their own list of OR journals as they were so unhappy with the ABS one) and tourism (Hall, 2011). The editors have responded several times (Morris, Harvey, Kelly and Rowlinson, 2011;Morris, Harvey and Kelly, 2009;Rowlinson, Harvey, Kelly and Morris, 2011). The main criticisms are:

i. The list has a limited coverage of journals – it included only 50% of the 1639 journals that were submitted to the 2008 RAE. Some fields had very few journals included. Because of the influence that the list acquired, journals that are not included tend to be discounted even if they are in fact of high quality.

ii. The panel of "experts" who did the rankings were not elected or representative of their communities and it was felt that there was bias against certain subjects. Much of the list is devoted to reference disciplines rather than to B&M itself and applied areas, for example economics alone accounts for 16% of the journals. Adding in psychology (5%) and social science (7%) covers 30% of the list. In contrast, in B&M we have general management (4%), HR (4%), marketing (7%), OR (4%) and strategy (2%). In total there are 22 different subject areas in the list – a very *ad hoc* selection as one of the editors accepts*:*

*"an eclectic mix of categories consisting of: academic disciplines, business functions, industries, sectors, issues or interests as well as more or less residual categories which includes many of the leading business and management journals"* (Rowlinson et al., 2013, p. 7)

iii. There was a huge discrepancy in the number of 4 grade journals across fields - psychology (42%), general management (23%), social science (20%), economics (13%), HR (11%), marketing (9%), finance (7%), ops management (3%), ethics/governance (0%), management education (0%) – again favouring the reference disciplines. This meant that in some fields it was actually impossible to get a paper in a 4 journal. In OR the only two 4 journals were both American and highly mathematical, refusing to publish papers in "soft OR" – a UK speciality.

iv. There was an over-reliance on ISI impact factors and other journal ranking lists. For example, many journals, some of very high quality, have never been included in the ISI Web of Science and so do not have impact factors. This downgrades them. Also, in later versions of the list, a new top category of 4* for "world elite" journals was introduced but this was largely based on appearances in other lists, such as the FT45 one, and so simply reproduces the status quo.

v. It was very difficult to generate change in the list. This particularly disadvantaged new journals that were not included, or journals that were improving their quality, or more application oriented journals. Because the list became so dominant, academics began to avoid publishing in journals that were not in ABS, preferably at least at the 3 level.

Whilst we have enumerated the shortcomings of the list, that is not our primary concern – any list has its particular biases and weaknesses. The real problem is that the rules of the REF, which required an evaluation of every output, and a selection of both outputs and staff, led to the ABS list coming to have a dominance over research in business and management that we consider to be extremely unhealthy.

The first problem is that quality of research in B&M has come to be identified almost exclusively with the quality level of the journal it is published in in ABS. "It's a 3* paper", "Jane Bloggs is a 4* researcher", "You need to get publishing in 3* journals" are now said, and taken for granted all the time. This ignores many things: that the range of qualities within a journal can be quite large as citation analyses show (Mingers, Macri and Petrovici, 2012a) with much of the quality overlapping; that there are many problems with the ABS quality levels as discussed above; that there are many high quality journals that are not included in the ABS list; and that it does not include books, book chapters etc.

Even worse, because of the REF, the emphasis is not just on ABS journals generally, but specifically the 3 and 4 level journals. The REF funding regime has increasingly focussed on the top end to the extent that first the 1* and 2* work was unfunded, and now there is little money available for even 3*. This has led the schools to focus, ever more determinedly, on 4* papers which have become the crown jewels of research publishing and what we might call 4* fetishism (Willmott, 2011). Academics themselves are not immune from this and the success of getting a 4* publication can become addictive.

The list has tended to focus, particularly at the top end, on well established, often US dominated, journals that tend to be theoretically oriented, very conservative in their scope and somewhat positivistic in the research approaches they accept. This itself can be traced to the original development of business schools in the US and their emulation of economics as the primary discipline in order to gain academic credibility (Khurana, 2007). We thus find ourselves in a situation in which there is a huge amount of pressure on researchers to get papers accepted into a fairly small number of mainstream 4 grade ABS journals (72 at the 4 level of which 22 are designated "world elite") which in turn become inundated with papers most of which have no chance of being accepted – the endless paper chase, or "Ring a ring of roses" as Macdonald and Kam (2007) put it. Research becomes ever-more traditional and mechanistic, turning the handle of the same few journals, rather than creative, innovative and driven by the pressing problems that face the organisational world (Tourish, 2011;Elton, 2000).

From a cultural capital perspective, the different RA strategies can be explained by the effects of cultural capital in opening the space for greater investment in the process of selectivity and having more to gain from the outcome of it for the striving universities. As Bourdieu argued, it is not just the deployment of cultural capital that is significant but the space for investment and reward it creates both going into the selectivity situation and emerging from it. Cultural capital in this case comprises the knowledge, skills, taste, preferences and possessions that give advantages (or disadvantages) in the system of relations.

We suggest that striving universities are particularly active in the selectivity game. More striving universities value league tables, shifting the nature of cultural capital from traditional outputs to journal papers. Conversely, fewer cultural capital resources for less striving universities means that they expect less out of the system and make less ambitious selectivity choices with less investment in those choices. These effects are compounded for less striving universities in two ways. The first is the converse of cultural capital, which is that less ambitious selectivity might be further conditioned by an unconscious anticipation of the limited resources at the branching points in the RA system. This would mean that less striving universities play safe by selecting very few staff rather than risking a more ambitious option. Furthermore there are constraints that bear down more forcefully on less striving universities. They lack the cultural capital that would expand the range of plausible possibilities of options and they have insufficient economic capital to sustain the risks of other choices, or more ambitious routes. Thus, universities themselves are an agent of elimination by favouring certain types of rules and by concealing selection decisions as neutral technical selection of the RA process.

The consequences of this REF-generated hegemony of the ABS list and the dominance of 3*/4* journals are multiple:

**The Suppression of Innovation**

Perhaps one of the most serious, we would argue, is the suppression of real innovative work, the kind of research that opens up whole new areas within a discipline, or across disciplines. The problem is the extreme focus on publishing in the ABS 3 and 4 journals. Such journals are usually very well established, very mainstream, with very traditional research approaches and long-standing, establishment figures as their editors and reviewers. They do not tend to publish speculative, blue-skies, boundary breaking research. Yet this is just what is needed to invigorate and enrich established research areas and, indeed, generate new ones. This kind of innovative work is often initiated by young researchers who have not yet been inculcated into the disciplinary matrix (Kuhn, 1977). Yet the forces outlined above inexorably channel such people into producing mechanistic and formulaic work to get their "four 3* papers" to assure their research careers. As Directors of Research, both the authors have found themselves advising young researchers "get the runs on the board" and leave the adventurous stuff till later. By which time, it is often too late as they have become thoroughly normalised.

We would like to illustrate this with two examples – one Professor Andrew Oswald's work on the economics of happiness as described in the *Times Higher Education* (Oswald, 2014) and the other our own work when we were ECRs. The study of happiness, from an economic perspective, is now well established (Frey, 2010) but in 1993 it had never been heard of. At that point, Oswald and a few colleagues decided to organise the first conference at LSE. "It is difficult to convey how strange, at that time, such an idea seemed. It lay somewhere between does-not-compute and ring-the-asylum" (p. 30). In the event there were only three people in the audience. However, this did not put them off, and they were not particularly concerned about whether or where their conference papers would get published. They were just interested in pursuing a stimulating idea. Had this been 2013 rather than 1993 it would almost certainly not have happened: "Unfortunately, I now witness a different set of attitudes among fellow academics … I see wonderful young scholars focussed on publishing per se and obsessed with satisfying the formal requirements of the research excellence framework" (p. 31). A conference on the same topic this year at LSE had an audience of 500 with many people turned away.

The second example concerns our own work in OR and management science in the 1980s. At that time, OR was almost exclusively a mathematical subject concerned with building models and simulations of organisational problems, what is now known as "Hard OR". However, a few people realised that working in organisations involved people, and people did not fit easily into mathematical models. This led us to explore work in sociology, psychology and philosophy such as critical theory, phenomenology and post-modernism in an effort to make OR more practically useful – "Soft OR". Such work was way beyond the boundaries of what was acceptable in the traditional, 3* and 4* journals such as *Management Science* and *Operations Research* and had to be published in niche journals such as the *Journal of Applied Systems Analysis* (now defunct) (Mingers, 1980, 1984) and *Omega* (then a 2* at best, now a 3*/4*) (White and Taket, 1996). Now, Soft OR and problem structuring methods (PSMs) have become fully established throughout the world except, interestingly enough, in the US where these two journals, the only 4* OR journals in the ABS list, still refuse to recognise non-mathematical OR (Mingers, 2011). This means that those of us working in Soft OR have no 4* journal to publish in, according to ABS. Ironically, this group includes the Chair of the Business and Management REF Panel, and two of its members. Hopefully, the Panel will live up to their promise not to slavishly follow the ABS list.

**Salami-Slicing Research Projects**

One of the effects of the emphasis on papers and the concomitant downgrading of books is that there is little motivation for researchers to undertake major research projects investigating really big questions. Such explorations may require several years before coming to fruition, involve a number of different researchers at multiple universities, would require a book to do justice to their results and probably serious grant money to undertake them. None of this fits in with the imperatives of the REF which requires several, top quality publications in the time it would take to do the research. The practical reality is that most academics cannot really afford to devote the necessary time to grant applications when they are highly risky and, even if successful, are treated by the REF panel as an "input" rather than an "output" generating a requirement for even more papers! Where significant research projects are undertaken, they tend to be structured into "bite-size" pieces that can generate papers along the way, each of which may be relatively weak. And there is little or no incentive to actually write books any more. One of the authors has written two research monographs within the period of the last two REFs neither of which were actually submitted!

There are also disincentives to collaborative work – while the same paper can be submitted by researchers from different institutions, it cannot generally within the same institution. This can cause problems in research teams who publish all their papers together, especially in terms of allocating them to individuals where they are of different quality, or indeed if there are not at least 4n papers between the n people.

**Disciplining Inter-Disciplinary Work**

As we said above, theoretically the REF wants to encourage inter-disciplinary work but the focus on 3*/4* journals goes right against this. Almost exclusively, the "top" (in an ABS sense) journals are strongly disciplinary, whether it is marketing, management science, accounting, finance or information systems. The chance of getting a truly inter- or trans-disciplinary paper accepted into these is quite remote. Even the so-called general management journals tend to have a fairly circumscribed theoretical and methodological approach. There are journals that will publish, and indeed seek, inter-disciplinary work (for example *Kybernetes*, *Systems Research and Behavioural Science*, *Systems Practice and Action Research*) but they tend to either not be in ABS, or be lowly rated.

A rigorous bibliometric study by Rafols et al (2012) demonstrated that journal rankings can disadvantage interdisciplinary work. They compared Innovation Studies units located within business schools in the UK and found that: i) Innovation Studies units were consistently more interdisciplinary than their B&M counterparts; ii) that the top journals in the ABS list covered a less diverse range of journals than did the lower ranked journals; and that therefore iii) B&M schools that were more disciplinary focussed were assessed more favourably. Their overall conclusion was that supposedly excellence-based journal ranking lists such as the ABS one systematically disfavour interdisciplinary work.

**Marginalising Practical Engagement**

Another effect of the ABS 4* fetishism is the marginalisation or indeed suppression of papers based on practical, engaged work with external organisations. This is all very ironic at a time when the REF is stressing impact by including it as a category in its own right. Even here, the research on which the impact is based has to be considered to be of at least 2* quality. Once again, the key problem is that the top ABS journals are all heavily theoretically oriented and do not easily publish case study based work unless it can be seen as making a major contribution to theory (Tourish, 2011). For example, one of the authors had a paper desk rejected (i.e., not even refereed) by both the *J. of Management Studies* and the *British Journal of Management* purely on the grounds that it was empirical rather than theoretical.

Part of the problem is that often this leads to a bifurcation of researchers into those oriented towards practice and those oriented towards theory with very few able to do both. Practical work is very time-consuming in terms of the process of engaging with external organisations, and tends to be driven by answering the problems of the organisation rather than addressing points of theory. Theoretical work is also very time-consuming in terms of a comprehensive and up-to-date knowledge of the literature, and an ability to address often rather abstract and arcane questions. The result can be, and indeed was at one of our institutions, that the practice-focussed researchers generated nearly all the impact case studies in the 2014 REF but were not submitted because their papers were not in top journals, while the majority of those submitted did not actually have any practical applications.

One might argue, so be it – some researchers do theoretical work, some do more practical work, and that might be reasonable if both were valued equally highly, but they are clearly not at least in terms of the REF and the ABS list. This is particularly a concern at the moment when there is a great deal of soul-searching in the business school world about their true purpose and their lack of engagement with the real, important problems that the world faces (Tranfield and Starkey, 1998;Starkey and Madan, 2001;Fincham and Clark, 2009;Hodgkinson and Rousseau, 2009;Kieser and Leiner, 2009;Syed, Mingers and Murray, 2009;van de Ven and Johnson, 2006;Khurana, 2007;Reed, 2009).

Here the debates surrounding around ranking fetishism appears to be closely related the institutionalised cultural capital (Bourdieu, 1986) which consists of institutional recognition, in the form of ranked journal articles, of the cultural capital held by an individual. This concept plays its most prominent role in the RA game, in which it allows a wide array of cultural capital to be expressed in a single measurement. The institutional recognition process thereby eases the conversion of cultural capital to other forms of capital by serving as a device that academics can use to describe their capital and universities can use to describe their needs for that capital.

With respect to the researchers' orientation, university business school researchers can possess capital that is practice in orientation or theory oriented capital and in which they are at the same time, desirable and undesirable, valued and disparaged, depending, possibly, on the striving nature of the university. In other words, we argue that there is a conceptualisation of cultural capital in terms of prestigious, theoretical pursuits, and an insistence that it can be conceptually distinguished from practical endeavours and this can be described as a dominant form.

**The Destruction of the Journal Ecosystem**

The pressures outlined above leading to such a focus on ABS 3* and 4* journals are also having a serious and detrimental effect on the whole population of journals themselves. There were over 1600 different journals submitted in the 2008 RAE (Mingers et al., 2012b) to the Business and Management panel, and this ignores those journals that were submitted to the Library and Information Science, and Accounting and Finance Panels which could also be seen as part of B&M. But in the 2011 ABS list only 94 are graded as 4 and a further 230 as 3. We have now reached a stage where, as far as the REF goes, and therefore as far as most business schools go, journals outside these select few are not worth publishing in. This poses significant problems for journals that are not in ABS, ranked as 1 or 2 in ABS, newly established journals, or niche journals.

Let us consider some particular examples. Stewart (2005) highlights the field of human resource development which is a new and emerging subject. There are several journals in the area – *Human Resource Development Quarterly, Human Resource Development Review* and *Human Resource Development International* – but they are all ranked as 2s in the 2010 ABS list. This means that academics in the area have to publish papers in other, more general, management journals if they are to be entered in the REF, and of course means that it is very difficult for these journals to get good papers and thus raise their ranking.

Consider also *Leadership* (Tourish, 2011). This was a new journal only founded in 2005. It had a mission to publish both theoretical and empirical work and also to encourage a wide range of theoretical and methodological perspectives. Fourteen papers were submitted in the 2008 RAE and analysis suggests that they may have been rated reasonably highly (Mingers et al., 2012b). However, *Leadership* is only graded as a 1 in the ABS list, primarily because it is new and has not had time, for example, to be included in the Web of Science, one of the main factors behind the ABS list.

Northcott and Linacre(2010) conducted a survey and interviews with authors, editors and publishers associated with accounting journals. Their conclusions were that:

> *"the entrenchment of NRAE [national RAE] 'rules' and journal quality perceptions has changed authors' submission choices and left lower ranked journals struggling with a diminished quantity and quality of submissions. A clear perception is that NRAEs have done little to improve the overall quality of the accounting literature, but are impeding the diversity, originality and practical relevance of accounting research"* (p. 38)

As well as damaging the diversity of publication outlets, the REF is diminishing the types of contributions that academics are prepared to make – if it does not count for the REF then they are not prepared to do it (Alldred and Miller, 2007). Journals (e.g., *Systems Research and Behavioural Science*) no longer publish book reviews (and there are fewer books to review); finding referees is increasingly difficult and yet there are more papers to review; and many of the other general academic tasks such as editorships, special issues and serving on professional bodies are seen as counter-productive.

**The Fragmentation of the Academic Community**

So far we have considered the effect of the RAE specifically on research, but in this section we want to widen the critique to include the effects on the academic community more generally which may, in the long run, be even more damaging.

Perhaps naively, we would argue that the academic community was, once, a community – a group of people interested in and committed to the discovery and transmission of knowledge. Whilst there were, of course, rivalries and disputes between research groups and disciplines there was common ground that research should be relatively unfettered, up to the interest and expertise of the individual; that academics should undertake a range of activities – teaching, research and administration; that departments should be relatively autonomous and self-governing; and that the highest values were integrity and innovative thinking. Successive research assessments have significantly changed this, shifting the balance from collegiality to managerialism (Yokoyama, 2006).

Perhaps the most obvious, and insidious, effect of the RAE has been an increasing split between those considered research active, because they are able to be submitted to an assessment, and the "rest" who are often forcibly put onto teaching-only contracts. It used to be the case that there was a range of research intensity from those who were genuinely only interested in teaching (and were often the best, particularly in an MBA context where practical experience is a must), through those who did a proportion of both, to the strong, excellent researchers (who were often very poor teachers!). Workloads could be adjusted to reflect individual strengths and staff might actually change the balance through their career. Moreover, as research intensive universities, not to mention QAA and HEFCE, argue, good teaching should be research led.

However, the drive towards submitting ever higher levels of research involving fewer and fewer staff, combined with the need not to appear highly selective has created a "success to the successful" archetype (Senge, 1990). Those staff capable of getting 3*/4* papers are given ever more resources – time in workload allocation models therefore less teaching, research funds and general approbation and are therefore able to be even more successful. Those who struggle for one reason or another, get less resource, more teaching and general opprobrium. They then have to become "teaching fellows" or indeed are encouraged to move elsewhere.

This can have significant negative effects on the individual – they become very stressed, they may perceive themselves as failures for not being in the research elite, and indeed their self-identity may change. There may be different kinds of change in different institutions. In the research intensive universities, where staff were always expected to research, the main problems will be for those deemed "not research active". However, in the new universities, which were traditionally teaching oriented, the pressure is on staff to start to carry out research even if this is not how they see themselves (Sikes, 2006). As well as problems within departments, the research assessment has also led to competition for staff between departments. As the stakes for doing well have become ever higher, the value of researchers with a good basket of 4* papers has become ever greater leading to a transfer market of staff movements, particularly in the run up to a REF submission (Piercy, 2000). Top researchers can command very significant salaries, and other inducements, to move to a competitor, and if their current institution wants to keep them they will have to match or exceed the offer – promotions, light teaching loads, large research funds and high salaries have become the norm. None of this actually improves overall research at all – there is only a limited supply – it just pushes up the price.

These pressures in the research domain also have their effects in the world of teaching which has itself become much more a concern in the time of high fees and the customerization of the student. In many universities teaching has now become devalued in comparison with research and it will be increasingly carried out by disgruntled staff who wanted to do research but have been forced on to teaching only contracts. Who can blame them for feeling like second class citizens when their research "colleagues" have study leaves, conference budgets, reduced teaching loads, and higher salaries? There will be a widening gap between teaching and research and it will no longer be possible to argue that teaching should be research led as who will be the researchers doing it? It will also widen once again the distance between the research intensive universities and those lower down who will become more teaching focused (Alldred and Miller, 2007).

Overall, the culture of academia has moved from one of collegiality to one of conflict and antagonism between researchers and teachers, and the managers (Deans and Directors of Research) and the managed. And this is not a recent occurrence – as early as 1997, at a Society for Research into Higher Education sponsored conference, Fulton in his summing up "particularly questioned whether the competitive, adversarial and punitive spirit evoked by the RAE was in the long run truly conducive to quality enhancement" (Elton, 2000, p. 279)

This can all be illustrated by one extreme example – the changes that have occurred at Warwick Business School (WBS) since a (relatively) poor result in the 2008 RAE, well documented by Parker (2014). WBS has always been one of the UK's leading business schools, seen as research intensive, innovative, and engaged with industry. It had been in the top 3 schools in all the assessments until 2008 when it appeared to be in ninth position in the league tables (actually joint fifth with five other schools). This was seen as a disaster by the University and a new Dean was quickly parachuted in with little or no consultation or discussion. He set about transforming WBS in order to prepare for the 2014 REF. Within a few years, a large proportion of the academics, even the best researchers, had moved to other institutions and the OR/MS group, one of the top two in the country, was decimated. These were replaced with staff, many from overseas, whose main quality was that they had 4* publications, especially in US journals. Their fit, or teaching contribution, did not matter that much (Piercy, 2000).

In the event, these measures did not help much – WBS came only twelfth on GPA, worse than before having been more selective in their staff submission. Interestingly, on their website[5] they claim to be ranked fifth, basing this on the GPA of research outputs only and thus ignoring impact and research environment, a measure that no other universities use.

# DISCUSSION

Our discussion of the RA process aims, through the lens of Bourdieu's theory, to illustrate that the repetitive quality of much organised life cannot be explained by a consequentialist rational actor model, but by the preconditions of choice. The persistence of practices lies in their taken-for-granted character and their reproduction in structures that are, to a degree, self-sustaining. Bourdieu argues that 'habitus' makes it possible to inhabit institutions, to draw on them practically, enacting their organising principles and thus reproducing them but at the same time allowing for revisions and transformation (Bourdieu, 1990).

Bourdieu's cultural reproduction theory points to several strategies with strong cultural influences on the actual mechanism of RA (via cultural capital) and reinforced by the institutional culture of universities in favour of more striving departments and universities. Rather than trying to explain persisting inequalities in the context of university funding and expansion, we think that this perspective puts in focus a concern with the maintenance of a hierarchical university structure. The perspective suggests that different positions in that hierarchical structure give different starting points and therefore different distances to travel in relation to the RA system. Avoidance of failure and processes of selection are central to both. Strategy and agency both play a role but are different across different universities and agency is with the striving group or department.

The influence of cultural capital makes the striving universities and departments more strategic in the RA field. It makes them more effective in selection as well as elaborating the possibilities of inclusion. They are therefore seen as more active players, with greater amounts of knowledge of the system and abilities to game the system as well as being more strategically long term in planning their research strategy overall.

Conversely the cultural capital perspective also assumes the rational anticipation and realistic estimation of constraints that makes the less striving universities less strategic in this sense. We would argue such practice indicates a powerful affective connection with the mass education markets (Gonzales, 2014).

Given the above observations it is more parsimonious to assume that there is a single rational strategy that is common to the more striving and less-striving universities and departments. Relative risk aversion is prior to, and more fundamental in, explaining these behavioural distinctions. This accounts for the added effort on the part of some striving universities that resulted in strategies and behaviours to gain a greater amount of the RA funding cake and to avoid demotion in the league tables. The analytical element that explains the observed differences in behaviour between more-striving and less-striving universities is the same rational appraisal of costs and benefits conditioned by the constraints created by their respective positions in a hierarchical structure. Assuming a single strategic rationality operating across universities focuses analytical attention solidly on the constraints facing different rational agents both within and outside the university system and how these factors bear down on RA strategy.

We suggest that there is an ongoing, interactive relationship between the institutional culture of the university system and individual departments that continues to affect RA decisions. This interactive relationship may affect ongoing decisions. This may lead to a number of perverse effects that has been highlighted. Such a relationship involves the behaviour of vice chancellors, deans and research directors in response to the incentives they face due to the design of the institutional structures and performance management of the RA system overall. What is required is a broader understanding of the institutional culture that results from the incentives created by the performance management framework and affects the ongoing relationships between the different actors within the university system. We explore the policy implications or what can be done in the next section.

To summarise, it is plausible to assume that culture can be a component in explaining differences in RA strategy. It is also plausible to assume that all consumers of the RA act rationally in seeking to maximise their expected and individualised returns. Such decision-making does not take place in isolation, however. The institutional culture of the university system has an impact on selection decisions and subsequent RA outcomes. The extent to which the broader policy or performance management framework may work to alleviate or accentuate the effects of institutional culture and/or the differential constraints identified in our analysis is therefore the question we now address.

**WHAT CAN BE DONE?**

Given the above analysis, it is tempting to join with others such as Harzing and Adler and suggest a complete moratorium on research assessment and a return to the "good old days" of unfettered and unmonitored research activity. However, the reality is that assessment is here to stay – people are already preparing for the next REF in 2020. So, we wish to take a more pragmatic line and make some recommendations for change that we believe will ameliorate some of the ridiculous catch-22's of the current system but still allow an equitable assessment of current research performance and a guide for future development. In doing this, we recognize that any measurement system will generate particular, strategic responses from those who are measured (Wouters, 2014).

The first recommendation is that, in many subjects, the RA should move away from peer review and the evaluation of individual research outputs to a bibliometric-based system subject to peer overview. This idea was examined in the UK before the 2008 RAE and rejected, but we believe that both the methods and the data for bibliometric review are now much more reliable. Moed (2007) and van Raan (2005) have argued for just such a combination of advanced metrics with transparent peer review. It is important that the bibliometrics used are not crude ones such as simple counts of citations or papers, or journal impact factors but more sophisticated ones that take into account the differences in citation practices across different disciplines. Examples are the crown indicator from the Centre for Science and Technology Studies (Moed, 2010a;Waltman et al., 2010;Leydesdorff and Opthof, 2011;Mingers and Lipitakis, 2013) which normalizes citation counts relative to citation levels in particular, pre-defined fields, or the more recent source-normalised metrics (Moed, 2010c;Waltman and van Eck, 2013;Zitt, 2010) such as SNIP that are normalised to the numbers of references in the papers that are the sources of citations. And it is then necessary to embed these within a peer review framework that can evaluate the substantive content of the work under review.

Abramo and D'Angelo (2011) compared informed peer review (including the UK RAE) with bibliometrics on the natural and formal sciences in Italy and concluded that bibliometrics were clearly superior across a range of criteria – accuracy, robustness, validity, functionality, time and cost. They recognized that there were problems in the social sciences and humanities where citation data is often not available.

It is clear from many studies that the basic citation data, provided by specialized databases such as Web of Science (WoS) or Scopus, varies in availability across the disciplines (Moed, 2005;Mingers and Lipitakis, 2010;Mingers and Lipitakis, 2013;Nederhof, 2006;van Leeuwen, 2006). There are two main reasons – in the social sciences, and especially the humanities, much research does not appear in the form of journal articles. It is often in monographs or book chapters, or it may be artifactual, e.g., software, poetry or the staging for a play, none of which are included in the citation databases. And, where it is in a journal, the proportion of journals included in databases may well be low. Mingers and Lipitakis (2013)found that in the outputs of three business schools WoS only included 50% of the journal papers and 20% of the total publications. Nederhof (2006), in a study of bibliometrics in social science and humanities, suggests that it is important to develop citation analysis methods beyond WoS citations. We would also suggest that further development of alternative metrics (altmetrics) based on the web such as downloads, twitter/facebook mentions and so on will be important (Kouters and Thelwall, 2014).

In terms of a revised research assessment at the moment, we think it would be necessary to have a flexible arrangement that differed between subjects. In the sciences the balance would be heavily towards bibliometrics while in the arts and humanities it would be predominantly peer review informed by whatever citation data was available. In the social sciences it might vary from one unit of assessment to another depending on the situation.

The second recommendation concerns the researchers who are eligible and selected for inclusion. To be eligible for submission by a particular university a researcher must not only be in post on a specific census date but must also have been employed by the institution for at least n months, on at least a p% contract where n might be 6 or 12, and p might be 40 or 50. This would ensure that the people submitted were people genuinely employed by, and making a contribution to the particular institution and mitigate the transfer market and hiring in of itinerant researchers.

We also believe that the distinction between research active and teaching academics should not form part of the RA. Essentially, all academics employed by an institution should be registered for the assessment and the numbers of staff made available and considered as part of the evaluation. They should also have any research outputs that they produced in the relevant period entered. This is obviously a major change, and under current arrangements would not be feasible as it would generate too much work for the panels, but with the move to bibliometrics it should be quite possible. All institutions now have institutional repositories where which contain details of all outputs although these would obviously have to be audited. This means that the basic data would already be available at least bibliographically in electronic format and the bibliometric calculations would be carried out by a computer, so it would not matter much what the volume was. This could in any case be controlled to some extent by varying the period being evaluated. In current metrics approaches such as SNIP (Moed, 2010b) a three year window is used. This might well be appropriate for the sciences, but a longer 5-year one could be used in social science and humanities where the gestation period for research is longer (Mingers, 2008).

These actions would have major benefits in removing the need for game playing about who to submit; moving less active researchers onto teaching contracts; and having to decide which outputs to enter. It would then lead on to down-grading the influence of journal lists such as ABS since there would be no decisions for departments to make. Everyone and everything would become part of the assessment. This in turn would hopefully mean that managers could get on with supporting and encouraging research itself, in order to get the best results, rather than spending their time in exercises to see how best to manipulate the system.

# Appendix 1 Changing Research Assessments

See Table 1 for a quantitative overview.

**1986 – Research Selectivity Exercise**. Organised by the University Grants Committee. 37 subject areas ("cost centres"). Universities had to submit five outputs (in total) per department! Also, they could submit a four page description of the research in the department. The results were scored on a four point scale from "below average" to "outstanding".

**1989** - **Research Selectivity Exercise.** Organised by the University Funding Council. 152 subject units, 70 peer review panels. Units asked to submit two outputs for all members of staff, together with information on PhD students and research grants. Results were scored on a 5-point scale relating to national and international excellence.

**1992 – Research Assessment Exercise (RAE).** Organised by Higher Education Funding Council for England (HEFCE). 72 units of assessment, 63 subpanels. Units asked to submit two research publications and two other forms of public output only for research active staff. 2,800 submissions. Same 5-point scale as 1989 but with a letter grade showing the proportion of staff submitted. (HEFCE, 1992;Taylor, 1995)

**1996** - **Research Assessment Exercise (RAE).** Organised by Higher Education Funding Council for England (HEFCE). 69 units of assessment (UoA) with 60 panels. Up to four outputs per academic. The 5-point scale had two extra levels added – 3a and 3b, and 5* at the top (HEFCE, 1996)

**2001** - **Research Assessment Exercise (RAE).** Organised by Higher Education Funding Council for England (HEFCE). The same UoAs and Panels as 1996. Umbrella panels were formed to ensure more consistency. Up to four outputs per academic but reductions for special circumstances. Staff who moved shortly before the census date could split their outputs. Same scale as 1996. (HEFCE, 2001). Funding reduced for the lower ranks (below 3A). A thorough review was conducted by Sir Gareth Roberts.

**2008** - **Research Assessment Exercise (RAE).** Organised by Higher Education Funding Council for England (HEFCE). 67 subpanels overseen by 15 main panels. Major change to assessment – each research output to be scored on a 5-point scale (0 – 4*) from "not research" to "world leading". Funding for top three later reduced to only 3* and 4*.

Bibliometrics not used although it had been suggested that they would be. Results in "quality profiles" as proportion of research judged to be in each category. Weighted sum of outputs, research environment and esteem. League tables produced based on the "grade point average". It was stated that after 2008 the RAE would be replaced by an assessment system that would be "mainly metrics based" (HEFCE, 2009, p. 6)

**2014** - **Research Excellence Framework (REF).** Organised by Higher Education Funding Council for England (HEFCE). Major change is to include an assessment of non-academic "impact" as 20% of the total. The use of bibliometrics largely dropped although citation data is made available to panels if they request it.

|  | 1986 | 1989 | 1992 | 1996 | 2001 | 2008 | 2014 |
|---|---|---|---|---|---|---|---|
| UoAs | 37 | 152 | 72 | 69 | 69 | 67 | 36 |
| Panels |  | 70 | 63 | 60 | 60 | 67 | 4 main panels with 36 subpanels |
| No. of submissions |  |  | 2783 | 2894 | 2598 | 2363 | 1911 |
| No. of outputs |  |  |  |  | 203743 | 215507 | 191232 |
| No. of staff |  |  | 43000 | 48072 | 48022 | 52401 | 52077 |
| Impact case studies |  |  |  |  |  |  | 6975 |
| Results scale | 4-points | 5-points concerning national/international excellence | 5-points concerning national/international excellence + letter grade | 7-point scale – 1, 2, 3a, 3b, 4, 5, 5* | 7-point scale – 1, 2, 3a, 3b, 4, 5, 5* | 5-point quality profile from "unclassified", "nationally recognised" to "world leading" | Same 4-point scale as 2008 but 20% of the final result is from non-academic "impact" |

Table 1 Summary of the main dimensions of the research assessments in the UK where available (variety of sources)

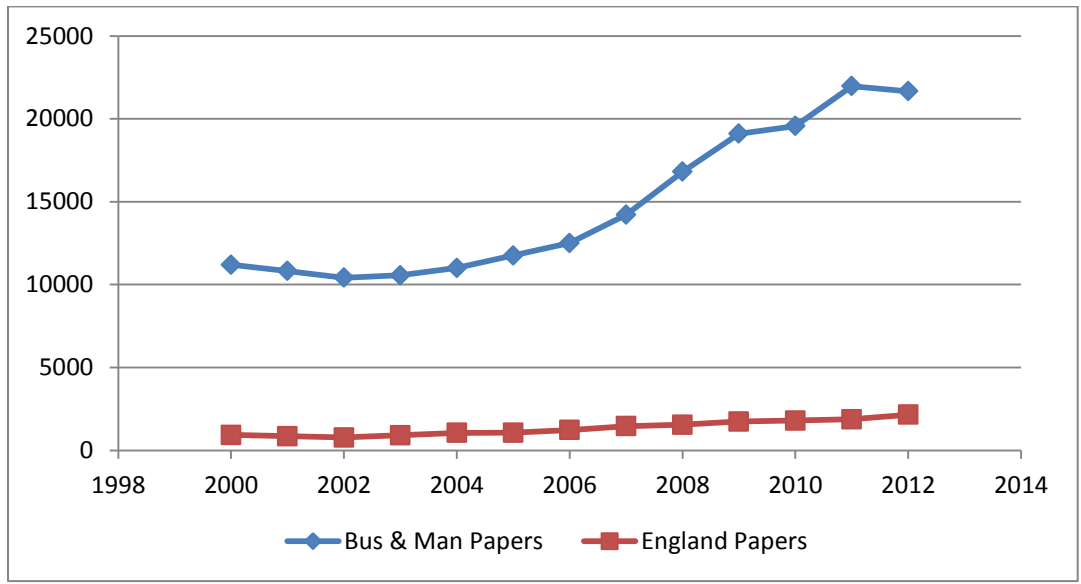

**Figure 1 Total number of articles in English in WoS categories "Business" and "Management", and the number where at least one author is from England**

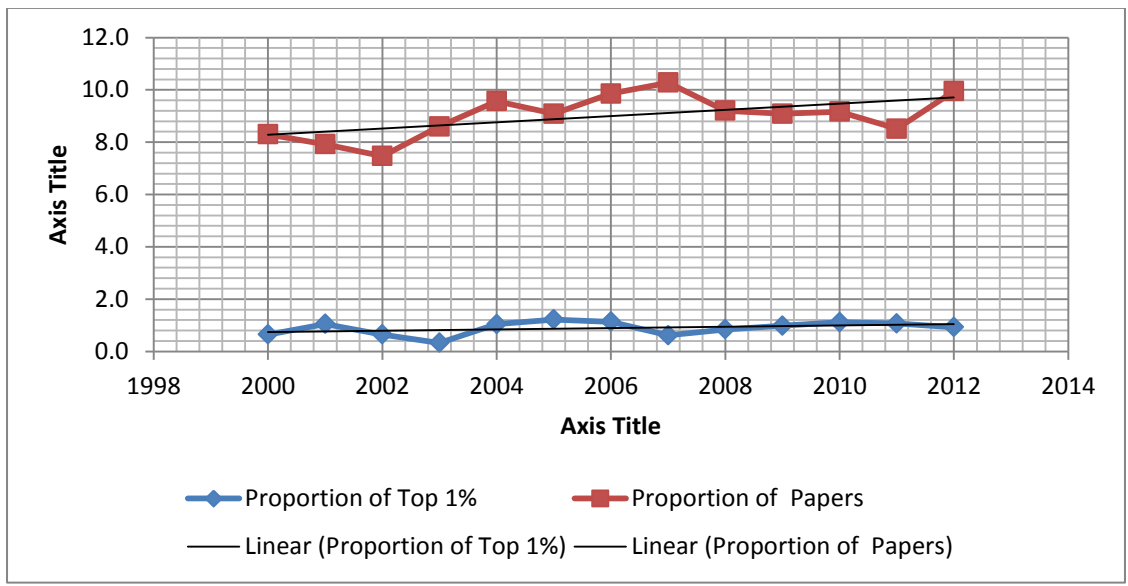

**Figure 2 Proportion of B&M papers from England (top line) and proportion of papers in the top 1* of citations**

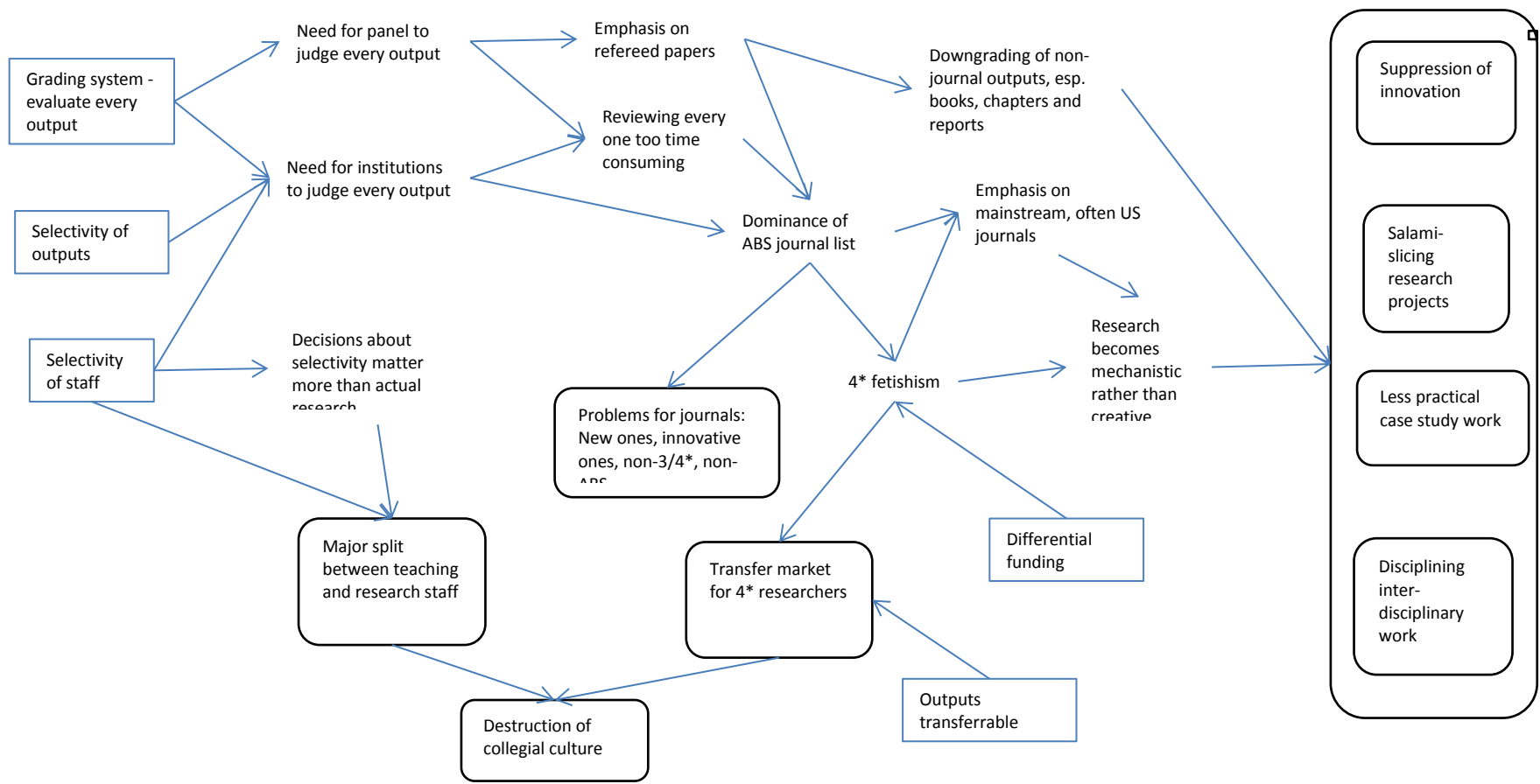

**Figure 3 Influence diagram of the c\uses and consequences of UK research assessment exercises.**

[1] This paper was written during Autumn 2014, before the REF results were published. When they did appear, they did not negate any of the comments made – if anything they exacerbated them. Some of the results are now incorporated in the paper.

[2] Over the years the name has changed several times. In this paper we shall use the generic name "research assessment" or RA unless we intend to refer to a specific one.

[3] This is partly to be expected as citations in science are much higher than social science, but nevertheless it is a very extreme result.

[4] Each Panel had to produce guidance notes about their process but as well, as that they held public meetings to explain more informally and to answer questions. These comments were not published, but the Panel Chair was very clear that refereed journal papers were the gold standard. When asked about books he indicated that they were harder to evaluate and it would be better to only submit those that had received good reviews in print, a form of refereeing. The same was true for research reports.

[5] http://www.wbs.ac.uk/news/wbs-ranked-fifth-for-research-output-and-third-for-research-power-in-the-uk-ref-2014/, accessed on 28th January 2015